\documentclass[aps,prd,groupedaddress,showpacs,nofootinbib]{revtex4}
\usepackage{graphicx,amssymb,amsmath}
\usepackage[english]{babel}

\newcommand{\be}{\begin{equation}}\newcommand{\ee}{\end{equation}}
\newcommand{\bea}{\begin{eqnarray}}\newcommand{\eea}{\end{eqnarray}}
\newcommand{\beaa}{\begin{eqnarray}}\newcommand{\eeaa}{\end{eqnarray}}
\newcommand{\ba}{\begin{array}}\newcommand{\ea}{\end{array}}
\newcommand{\bit}{\begin{itemize}}\newcommand{\eit}{\end{itemize}}
\newcommand{\ben}{\begin{enumerate}}\newcommand{\een}{\end{enumerate}}

\def\lab{\label}
\def\lan{\langle}
\def\lf{\left}

\def\ran{\rangle}
\def\rar{\rightarrow}
\def\ri{\right}

\def\al{\alpha}\def\bt{\beta}

\def\te{\theta}

\def\si{\sigma}
\def\om{\omega}

\def\1{{_{1}}}\def\2{{_{2}}}
\def\nof{:\;\!\!\;\!\!:}

\begin{document}

\title{Neutrino mixing, flavor states and dark energy}

\author{M.Blasone${}^{\flat}$, A.Capolupo${}^{\natural}$, S.Capozziello${}^{\sharp}$, G.Vitiello${}^{\flat}$}


\affiliation{${}^{\flat}$ Dipartimento di Matematica e
Informatica,
 Universit\`a di Salerno and INFN,
 Gruppo Collegato di Salerno, 84100 Salerno, Italy,
\\ ${}^{\natural}$ Department of Physics and Astronomy,
University of Leeds, Leeds LS2 9JT UK,
\\  ${}^{\sharp}$ Dipartimento di Scienze Fisiche, Universit\`a di Napoli "Federico II" and INFN Sez. di Napoli,
Compl. Univ. Monte S. Angelo, Ed.N, Via Cinthia, I-80126 Napoli,
Italy.}


\vspace{2mm}

\begin{abstract}

 We shortly summarize the quantum field theory formalism for the
neutrino mixing and report on recent results showing that the
vacuum condensate induced by neutrino mixing can be interpreted as
a dark energy component of the Universe.

\end{abstract}


\maketitle
The study of neutrino mixing
\cite{Pontecorvo:1957cp,Wolfenstein:1977ue} in the quantum field
theory (QFT) framework \cite{BV95}-\cite{Blasone:2006jx} has led to
the discovery of the unitary inequivalence between the massive
neutrino vacuum and the flavor vacuum. This has shed new light on
the proper definition of neutrino flavor states and has brought to a
deeper understanding of the non-perturbative nature of the particle
mixing phenomenon. Pontecorvo oscillation formulas have been
recognized to be the quantum mechanical approximation of the exact
QFT oscillation formulas and the vacuum energy due to the
neutrino-antineutrino pair condensate has been computed
\cite{Blasone:2004yh}, which can be interpreted as dynamically
evolving dark energy \cite{Capolupo:2006et}.

Here we summarize these results by considering two flavor neutrino
mixing. Extension to three flavors has been also done
\cite{BV95}-\cite{Capolupo:2006et}. The Pontecorvo mixing
transformations for Dirac neutrino fields are
\begin{eqnarray} \label{mix1}
\nu_{e}(x) &=&\nu_{1}(x)\,\cos\theta + \nu_{2}(x)\,\sin\theta
\\ [2mm]
\label{mix2} \nu_{\mu}(x) &=&-\nu_{1}(x)\,\sin\theta +
\nu_{2}(x)\,\cos\theta \;,\end{eqnarray}
where $\theta$ is the mixing angle, $\nu_{e}$ and $\nu_{\mu}$ are
the fields with definite flavors  and $\nu_1$ and $\nu_2$ are those
with definite masses $m_{1} \neq m_{2}$:
\begin{equation} \label{field}
 \nu _{i}(x)=\frac{1}{\sqrt{V}}{\sum_{{\bf k} ,
r}} \left[ u^{r}_{{\bf k},i}\, \al^{r}_{{\bf k},i}(t) + v^{r}_{-{\bf
k},i}\, \bt^{r\dag}_{-{\bf k},i}(t) \ri] e^{i {\bf k}\cdot{\bf x}}\,
,  \end{equation}
with  $i=1,2~, ~ r=1,2$, $ \al_{{\bf k},i}^{r}(t)=\al_{{\bf
k},i}^{r}\, e^{-i\omega _{k,i}t}$, $ \bt_{{\bf k},i}^{r\dag}(t) =
\bt_{{\bf k},i}^{r\dag}\, e^{i\omega_{k,i}t},$ and $ \omega
_{k,i}=\sqrt{{\bf k}^{2} + m_{i}^{2}}.$ The operators $\alpha
^{r}_{{\bf k},i}$ and $ \beta ^{r }_{{\bf k},i}$ annihilate the
vacuum state $|0\rangle_{1,2}\equiv|0\rangle_{1}\otimes
|0\rangle_{2}$.

The mixing transformation Eqs.(\ref{mix1}), (\ref{mix2}) can be
written as \cite{BV95}:
\bea \label{mixG} \nu_{e}^{\alpha}(x) = G^{-1}_{\bf \te}(t)\;
\nu_{1}^{\alpha}(x)\; G_{\bf \te}(t)
\\
 \nu_{\mu}^{\alpha}(x)
= G^{-1}_{\bf \te}(t)\; \nu_{2}^{\alpha}(x)\; G_{\bf \te}(t) \eea
where $G_{\bf \te}(t)$ is the mixing generator given by
\bea\label{generator12} G_{\bf \te}(t) = exp\left[\theta \int
d^{3}{\bf x} \left(\nu_{1}^{\dag}(x) \nu_{2}(x) - \nu_{2}^{\dag}(x)
\nu_{1}(x) \right)\right]\;. \eea
At finite volume, $G_{\bf \te}(t)$ is an unitary operator,
$G^{-1}_{\bf \te}(t)=G_{\bf -\te}(t)=G^{\dag}_{\bf \te}(t)$,
preserving the canonical anticommutation relations; it maps the
Hilbert space ${\cal H}_{1,2}$ for $\nu_1$ and $\nu_2$ fields  to
the Hilbert space ${\cal H}_{e,\mu}$ for the flavor fields. In
particular, $ |0(t) \rangle_{e,\mu} = G^{-1}_{\bf \te}(t)\; |0
\rangle_{1,2}\; $, where $|0 \rangle_{e,\mu} \in {\cal H}_{e,\mu}$
is the flavor vacuum. The crucial point is that in the infinite
volume limit ${\cal H}_{1,2}$ turns out to be unitarily inequivalent
to ${\cal H}_{e,\mu}$ \cite{BV95}. A feature which marks the
essential difference with the quantum mechanical formalism
\cite{BV95}.

The flavor annihilators, relative to the fields $\nu_{e}(x)$ and
$\nu_{\mu}(x)$, are $ \alpha _{{\bf k},\sigma}^{r}(t) \equiv
G^{-1}_{\bf \te}(t)\;\alpha _{{\bf k},i}^{r}(t)\;G_{\bf \te}(t), $
and similar for $\beta _{{\bf k},\sigma}^{r}$, with
$(\sigma,i)=(e,1), (\mu,2)$. In the reference frame such that ${\bf
k}=(0,0,|{\bf k}|)$ they are explicitly given by:
\bea\label{annihilator}  \alpha^{r}_{{\bf k},e}(t)=
\cos\theta\;\alpha^{r}_{{\bf k},1}(t)\;
 +\;\sin\theta\;\left( |U_{{\bf k}}|\; \alpha^{r}_{{\bf
k},2}(t)\;+\;\epsilon^{r}\; |V_{{\bf k}}|\; \beta^{r\dag}_{-{\bf
k},2}(t)\right) \eea
and similar for $\alpha^{r}_{{\bf k},\mu} $, $\beta^{r}_{{\bf
k},e}$, $ \beta^{r}_{{\bf k},\mu}$, where $ |U_{{\bf k}}| \equiv
u^{r\dag}_{{\bf k},i} u^{r}_{{\bf k},j} = v^{r\dag}_{-{\bf k},i}
v^{r}_{-{\bf k},j}\,, ~i,j = 1,2~, i \neq j~, \quad  |V_{{\bf k}}|
\equiv  \epsilon^{r}\; u^{r\dag}_{{\bf k},1} v^{r}_{-{\bf k},2} =
-\epsilon^{r}\; u^{r\dag}_{{\bf k},2} v^{r}_{-{\bf k},1} $. We have
$ |U_{{\bf k}}|^{2}+|V_{{\bf k}}|^{2}=1$. For the explicit
expressions of $U_{{\bf k}}$ and $V_{{\bf k}}$ see Ref. \cite{BV95}.

The number of condensed neutrinos in the flavor vacuum is given by
\bea \label{density} _{e,\mu}\langle 0| \al_{{\bf k},i}^{r \dag}
\al^r_{{\bf k},i} |0\rangle_{e,\mu} = {}_{e,\mu}\langle 0| \bt_{{\bf
k},i}^{r \dag} \bt^r_{{\bf k},i} |0\rangle_{e,\mu} = \sin^{2}\te
|V_{{\bf k}}|^{2},
 \eea
with $i=1,2$. Notice that $|V_{{\bf k}}|^{2} = 0$ for $m_1 = m_2$,
it has a maximum at $|{\bf k}|^2 = m_{1} m_{2}$ and goes to zero for
large momenta (i.e. for $|{\bf k}|^2\gg m_{1} m_{2}$ ) as $|V_{{\bf
k}}|^2 \approx \frac{(\Delta m)^2}{4 k^2}$. As we will see below,
the mixing of neutrinos contributes to the dark energy exactly
because of the non-zero value of $|V_{\bf k}|^2$
\cite{Blasone:2004yh}.

Flavor neutrinos are produced in charged current weak interaction
processes, like $ W^{+} \rightarrow e^{+} + \nu_{e}$, as described
in the Standard Model. At tree level, flavor charge is strictly
conserved in the vertex; field mixing allows for the possibility of
violation of lepton number via loop corrections. Such corrections
are however extremely small and practically unobservable. In
practice neutrinos are identified by the observation of the
corresponding charged leptons, {\it assuming} flavor conservation in
production (detection) vertices. In the presence of the neutrino
mixing, however, one has to consider the fact that selecting to work
with the Hilbert space of the mass eigenstates poses a mathematical
problem, because of the existence of a separate Hilbert space (the
one for the flavor eigenstates), as well as a physical problem since
it amounts to work with the Pontecorvo states which are not
eigenstates of the flavor charges, in contrast with the observed
conservation of lepton number in the neutrino production
(detection).  Lepton numbers are indeed good quantum numbers,
provided the neutrino production (detection) vertex can be localized
within a region much smaller than the region where flavor
oscillations take place, which is what happens in practice, since
typically the spatial extension of the neutrino source (detector) is
much smaller than the neutrino oscillation length.

Thus, flavor neutrino states have to be eigenstates of the neutrino
flavor charges $Q_{\nu_e}$ and $ Q_{\nu_\mu}$. These operators may
be expressed in terms of the (conserved) charges for the neutrinos
with definite masses $Q_{\nu_1}$ and $Q_{\nu_2}$ as follows
\cite{Blasone:2006jx}:
\bea Q_{\nu_e}(t) &=& \cos^2\te\;
Q_{\nu_1} + \sin^2\te \; Q_{\nu_2}  +\sin\te\cos\te \int d^3{\bf x} \lf[\nu_1^\dag (x) \nu_2(x) +
\nu_2^\dag(x) \nu_1(x)\ri]\,, \label{carichemix1}
\\
 Q_{\nu_\mu}(t) &=& \sin^{2}\te \;
Q_{\nu_1} +\cos^{2}\te \; Q_{\nu_2}  -\sin\te \cos\te \int d^3{\bf x} \lf[\nu_1^\dag(x) \nu_2(x) +
\nu_2^\dag(x) \nu_1(x)\ri]\,. \label{carichemix2}\eea
We remark that the last term in the above expressions forbids the
construction of eigenstates of the $Q_{\nu_\sigma}(t)$,
$\sigma=e,\mu$, in the Hilbert space
 ${\cal H}_{1,2}$ for the fields with definite masses.
One can also show that the above flavor charge operators are
diagonal in the ladder operators $\al_{\si}$, $ \beta_{\si}$, for
the neutrino flavor fields:
\bea \nof Q_{\nu_\sigma}(t) \nof \; \,= \,\;
\sum_{r} \int d^3 {\bf k} \, \lf( \al^{r\dag}_{{\bf k},\si}(t)
\al^{r}_{{\bf k},\si}(t)\, -\, \beta^{r\dag}_{-{\bf k},\si}(t)
\beta^{r}_{-{\bf k},\si}(t)\ri)\, , \label{flavchadiag} \eea
where $\sigma=e,\mu$ and  $\nof ... \nof\,$ denotes normal ordering
with respect to $|0\ran_{e,\mu}$. This makes straightforward the
definition (at reference time $t=0$) of  flavor  neutrino and
antineutrino states as:
\bea\label{flavstate} |\nu^{r}_{\;{\bf k},\si}\ran \equiv
\al^{r\dag}_{{\bf k},{\sigma}} |0\ran_{{e,\mu}}~; \quad |{\bar
\nu}^{r}_{\;{\bf k},\si}\ran \equiv \bt^{r\dag}_{{\bf k},{\sigma}}
|0\ran_{{e,\mu}}\, , ~~\si = e,\mu ~, \eea
which are thus by construction eigenstates of the operators  in
Eq.(\ref{flavchadiag}) at $t=0$, as it should be in agreement with
the observed conservation of flavor charge in the neutrino
production (detection) vertices \cite{Blasone:2006jx}.

The oscillation formulas are  obtained by taking expectation values
of the above charges on the flavor neutrino states. Consider for
example an initial electron neutrino. Working in the Heisenberg
picture, we obtain
\bea {\cal Q}^{\bf k}_{\nu_\si}(t)  \,\equiv\, \langle
\nu^{r}_{{\bf k},e}|Q_{\nu_{\si}}(t)
|\nu^{r}_{\;{\bf k},e}\ran \; =\, \lf|\lf \{\al^{r}_{{\bf k},\si}(t), \al^{r \dag}_{{\bf
k},e}(0) \ri\}\ri|^{2} \;+ \;\lf|\lf\{\bt_{{-\bf k},\si}^{r
\dag}(t), \al^{r \dag}_{{\bf k},e}(0) \ri\}\ri|^{2} \, .
\label{charge1}\eea

Charge conservation is obviously ensured at any time: ${\cal Q}^{\bf k}_{\nu_e}(t)
+ {\cal Q}^{\bf k}_{\nu_\mu}(t)\; = \; 1$ and
$_{e,\mu}\langle 0|Q_{\nu_{\si}}(t)| 0\rangle_{e,\mu} =  0 $.
The oscillation (in time) formula for the flavor charge on an
initially electron neutrino state, is thus:
\bea \label{enumber} {\cal Q}^{\bf k}_{\nu_e}(t)   \, = \,1 - \sin^{2}( 2
\te) \, |U_{{\bf k}}|^{2}\; \sin^{2} \lf( \frac{\om_{k,2} -
\om_{k,1}}{2} t \ri)
- \sin^{2}( 2 \te)\,|V_{{\bf k}}|^{2} \; \sin^{2} \lf(
\frac{\om_{k,2} + \om_{k,1}}{2} t \ri) \, ,
\end{eqnarray}
%
with $|U_{{\bf k}}|^{2}+|V_{{\bf k}}|^{2}=1$. The differences with
respect to the usual formula for neutrino oscillations are in the
energy dependence of the amplitudes and in the additional
oscillating term. In the relativistic limit $|{\bf
k}|\gg\sqrt{m_1m_2}$, we have $|U_{{\bf k}}|^{2}\rar 1$ and
$|V_{{\bf k}}|^{2}\rar 0$ and the traditional (Pontecorvo)
oscillation formula is recovered \cite{BV95}.

As already observed,  contrarily to the QFT flavor states considered
above, the quantum mechanical (Pontecorvo) flavor states
\cite{Pontecorvo:1957cp}:
\begin{eqnarray} \label{nue0a}
|\nu^{r}_{\;{\bf k},e}\rangle_P &=& \cos\theta\;|\nu^{r}_{\;{\bf
k},1}\rangle \;+\; \sin\theta\; |\nu^{r}_{\;{\bf k},2}\rangle \,,
\\ [2mm] \label{nue0b}
|\nu^{r}_{\;{\bf k},\mu}\rangle_P &=& -\sin\theta\;|\nu^{r}_{\;{\bf
k},1}\rangle \;+\; \cos\theta\; |\nu^{r}_{\;{\bf k},2}\rangle \, ,
\end{eqnarray}
are {\em not} eigenstates of the flavor charges
\cite{Blasone:2005ae}. This can be explicitly seen by using
Eqs.(\ref{carichemix1}) and (\ref{carichemix2}). One can estimate
how much the flavor charge is violated in the usual quantum
mechanical states by taking the expectation values of the flavor
charges on the above Pontecorvo states. Considering for example an
electron neutrino state, we obtain:
\bea  _{P}\langle\nu^{r}_{{\bf k},e}|: Q_{\nu_e}:
|\nu^{r}_{\,{\bf k},e}\rangle_{P} &=& \cos ^{4}\theta +
\sin^{4}\theta +  2 |U_{\bf k}| \sin^{2}\theta \cos^{2}\theta < 1, \label{caric} \\
[2mm] {}_{P}\langle\nu^{r}_{{\bf k},e}| :Q_{\nu_\mu}: |\nu^{r}_{{\bf
k},e}\rangle_{P} &=& 2 (1 - |U_{\bf k}|) \sin^{2}\theta
\cos^{2}\theta  \,>0, \eea for any $ \theta \neq 0,  m_{1} \neq
m_{2},\; {\bf k} \neq 0$ and where  $: ... :\,$ denotes normal
ordering with respect to the vacuum state $|0\ran_{1,2}$. In the
relativistic limit, $|U_{\bf k}| \rar 1$ \cite{BV95} and the
Pontecorvo states are a good approximation for the exact charge
eigenstates Eq.(\ref{flavstate}).

We now turn our attention to the vacuum energy due to neutrino
condensate. We calculate the contribution $\rho_{vac}^{mix}$ of the
neutrino mixing to the vacuum energy density in a Minkowski
space-time, but the computation can be extended to curved
space-times. At the present epoch it is a good approximation of that
in FRW space-time, since  the characteristic
oscillation length of the neutrino is much smaller than the universe
curvature radius. Computations carried on in a curved background
give a time dependent dark energy, leading, however, to the same
final result we obtain in the flat space-time.

The Lorentz invariance of the vacuum requires that the vacuum
energy-momentum tensor density ${\cal T}_{\mu\nu}^{vac}$ is equal to
zero.
 The (0,0) component of ${\cal
T}_{\mu\nu}(x) $ for the fields $\nu_1$ and $\nu_2$ is $ :{\cal
T}_{00}(x): ~= \frac{i}{2}:\left({\bar \Psi}_{m}(x)\gamma_{0}
\stackrel{\leftrightarrow}{\partial}_{0} \Psi_{m}(x)\right): $,
where $\Psi_{m} = (\nu_1, \nu_2)^{T}$. The (0,0) component of the energy-momentum tensor $
T_{00}=\int d^{3}x {\cal T}_{00}(x)$ is then given by $\,
:T^{00}_{(i)}: ~= \sum_{r}\int d^{3}{\bf k}\,
\omega_{k,i}\lf(\al_{{\bf k},i}^{r\dag} \al_{{\bf k},i}^{r}+
\beta_{{\bf -k},i}^{r\dag}\beta_{{\bf -k},i}^{r}\ri),$ with $i=1,2$.
We note that $T^{00}_{(i)}$ is time independent.

In the early universe epochs, when the Lorentz invariance of the
vacuum condensate is broken,
 $\rho_{vac}^{mix}$ presents also space-time dependent condensate
 contributions \cite{Capolupo:2006et}.
Then $\rho_{vac}^{mix}$ and the contribution of the neutrino  mixing
to the vacuum pressure
 $ p_{vac}^{mix}$ are given respectively by:
 \bea
\rho_{vac}^{mix}= -\frac{1}{V}\; \eta_{00} \; {}_{e,\mu}\lan 0
|\sum_{i} :T^{00}_{(i)}:| 0\ran_{e,\mu}\,,
\\
p_{vac}^{mix}= -\frac{1}{V}\; \eta_{jj} \; {}_{e,\mu}\lan 0
|\sum_{i} :T^{jj}_{(i)}:| 0\ran_{e,\mu}\,,
 \eea
where $ :T^{jj}_{(i)}:= \sum_{r}\int d^{3}{\bf k}\, \frac{k^j
k^j}{\;\omega_{k,i}}\lf(\al_{{\bf k},i}^{r\dag} \al_{{\bf k},i}^{r}+
\beta_{{\bf -k},i}^{r\dag}\beta_{{\bf -k},i}^{r}\ri)$ (no summation
on the index $j$ is intended). We then obtain
\bea   \rho_{vac}^{mix} &=& \sum_{i,r}\int
\frac{d^{3}{\bf k}}{(2 \pi)^{3}} \omega_{k,i}
 \Big({}_{e,\mu}\lan 0 |\al_{{\bf k},i}^{r\dag} \al_{{\bf
k},i}^{r}| 0\ran_{e,\mu} + {}_{e,\mu}\lan 0 |\beta_{{\bf
k},i}^{r\dag} \beta_{{\bf k},i}^{r}| 0\ran_{e,\mu} \Big) ,  \eea
and, by introducing the cut-off $K$,
\bea\label{cc}
 \rho_{vac}^{mix} = \frac{ 2}{\pi} \sin^{2}\theta
\int_{0}^{K} dk \, k^{2}(\omega_{k,1}+\omega_{k,2}) |V_{\bf k}|^{2}
\,. \eea

We also obtain
 \bea\label{cc2}
  p_{vac}^{mix} = \frac{2}{3\;\pi}
\sin^{2}\theta \int_{0}^{K} dk \,
k^{4}\lf[\frac{1}{\omega_{k,1}}+\frac{1}{\omega_{k,2}}\ri] |V_{\bf
k}|^{2}\,.
 \eea

Eqs.(\ref{cc}) and (\ref{cc2}) show that the adiabatic index is $w =
p_{vac}^{mix}/ \rho_{vac}^{mix} \simeq 1/3$ when the cut-off is
chosen to be $K \gg m_{1}, m_{2} $ \cite{Capolupo:2006et}.

The values of $\rho_{vac}^{mix}$ and  $ p_{vac}^{mix}$ which we
obtain are time-independent since we are taking into account the
Minkowski metric. In a curved space-time, time-dependence has to be
taken into account, but the essence of the result is the same.

At the present epoch, the breaking of the Lorentz invariance is
negligible and then the contribution
 $\rho^{mix}_{\Lambda}$ of the neutrino mixing to the vacuum energy density
  comes from space-time independent condensate contributions
 (i.e. the contributions carrying a non-vanishing $\partial_{\mu} \sim k_{\mu}=(\omega_{k},k_{j}) $
are missing).
 Thus, the energy-momentum density tensor of the vacuum condensate
is
\bea \lab{tmunu}
  {}_{e,\mu}\lan 0 |:T_{\mu\nu}:| 0\ran_{e,\mu} =
\eta_{\mu\nu} \sum_{i}m_{i}\int
\frac{d^{3}x}{(2\pi)^3}\;{}_{e,\mu}\lan 0 |:\bar{\nu
}_{i}(x)\nu_{i}(x):| 0\ran_{e,\mu} = \eta_{\mu\nu}
\rho_{\Lambda}^{mix}.  \eea
Since $\eta_{\mu\nu} = diag (1,-1,-1,-1)$ and, in a homogeneous and
isotropic universe,  $T_{\mu\nu} = diag (\rho\,,p\,,p\,,p\,)$, then
the state equation is $\rho_{\Lambda}^{mix} = -p_{\Lambda}^{mix}$.
This means that $\rho_{\Lambda}^{mix}$ contributes today to the
dynamics of the universe by a
 cosmological constant behavior \cite{Capolupo:2006et}. We
obtain
\bea\label{cost}
\rho_{\Lambda}^{mix}= \frac{2}{\pi}
\sin^{2}\theta \int_{0}^{K} dk \,
k^{2}\lf[\frac{m_{1}^{2}}{\omega_{k,1}}+\frac{m_{2}^{2}}
{\omega_{k,2}}\ri] |V_{\bf k}|^{2}.
\eea
 Notice that
 $\rho_{\Lambda}^{mix}$ would be zero for $|V_{\bf k}|^{2}= 0$ for
any $|\bf k|$, as it is in the usual (Pontecorvo) approximation.

One can show \cite{Capolupo:2006et} that the integral in Eq.
(\ref{cost}) diverges in $K$ as $m_{i}^{4}\,\log\lf( 2K/m_{j} \ri)$.
However, the divergence in $K$ is smoothed by the factor
$m_{i}^{4}$. For neutrino masses of order of $10^{-3} eV$ and
$\Delta m_{12}^{2} \approx 10^{-5} eV^{2}$ we have
$\rho_{\Lambda}^{mix} \approx 2.9 \times 10^{-47} GeV^{4}$, which is
the observed dark energy value, even for a value of the cut-off of
order of the Planck scale $K=10^{19} GeV$ (see Fig. \ref{Fig1}). One
can also see that $\frac{d \rho_{\Lambda}^{mix}(K)}{d K} \propto
\frac{1}{K} \rightarrow 0$ for large $K$.

\begin{figure}
\centering \resizebox{9cm}{!}{\includegraphics{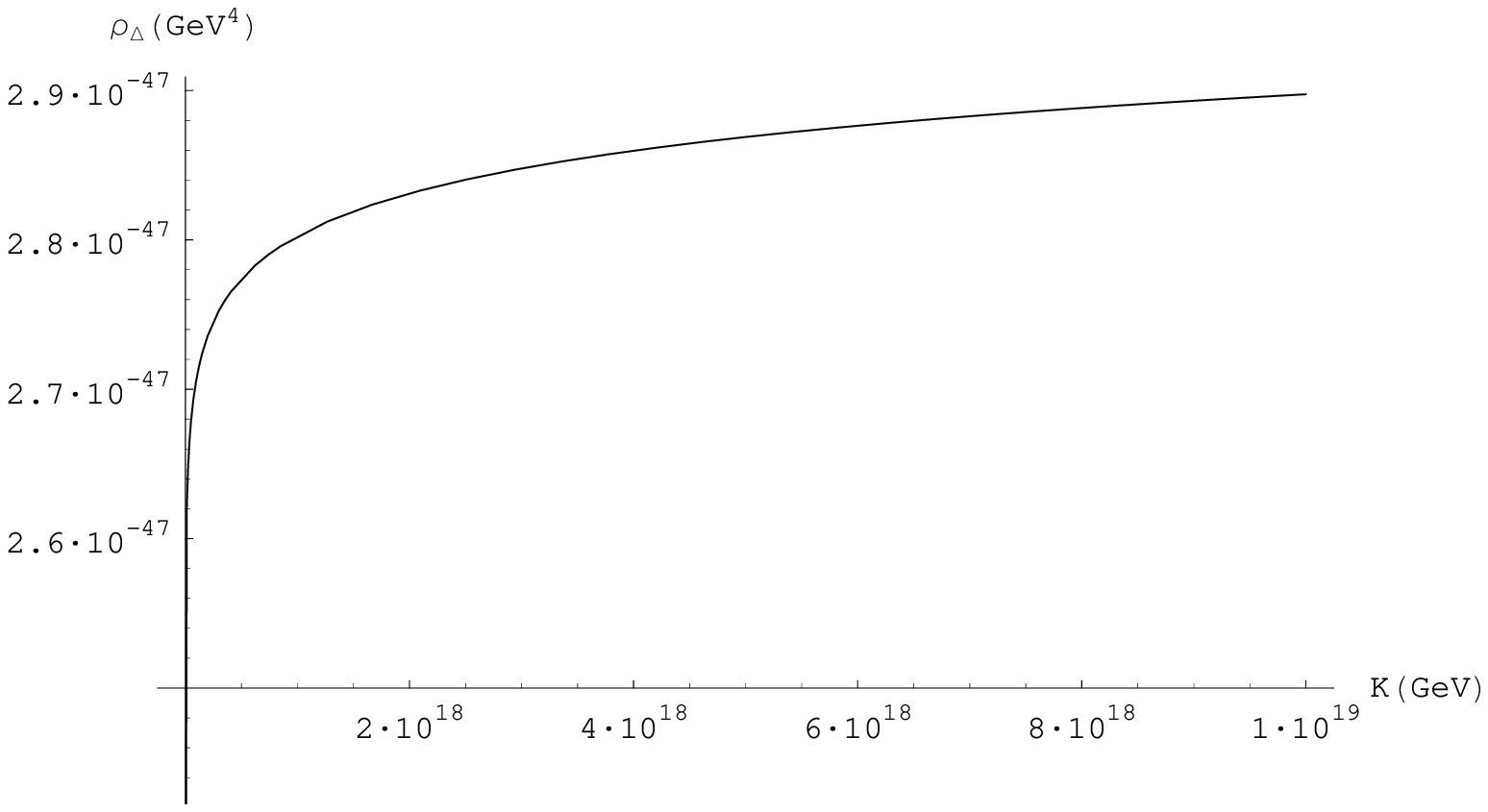}}
\hfill \caption{Plot of Eq. (\ref{cost}).} \label{Fig.1.}
\end{figure}

In conclusion, the vacuum condensate generated by neutrino mixing
can be interpreted as an evolving dark energy that, at present
epoch, behaves as the cosmological constant.

When the three generation mixing is considered we obtain a value for
$\rho_{\Lambda}^{mix}$ which differs of about $5$ order of magnitude
from the observed dark energy value. It has been proposed in the
literature \cite{Mavromatos:2007sp} that heaviest neutrinos should
not contribute to the vacuum energy density. The present result
might suggest that this is indeed the case. However, it also points
to the need of a further theoretical refinement of the present
approach, which is in our future research plans. At present state of
art, among the merits of the complete QFT treatment of neutrino
mixing above summarized there is the remarkable improvement in the
computed order of magnitude of the dark energy (to be compared with
the $123$ order of magnitude of disagreement in usual approaches)
and the fact that in the present approach there is no need of
postulating (till now) unobserved exotic fields.

\end{document}